\documentclass[fleqn,twoside]{article}
\usepackage{espcrc2}
\usepackage{amsmath}
\usepackage{graphicx}

\newcommand{\dd}{{\mathrm{d}}}
\newcommand{\DD}{{\mathrm{D}}}
\newcommand{\ee}{{\mathrm{e}}}
\newcommand{\ii}{{\mathrm{i}}}

\newcommand{\order}{{\mathcal{O}}}
\newcommand{\q}{{\textrm{q}}}

\newcommand{\fluct}{{\textrm{fluct}}}
\newcommand{\tmin}{{\textrm{min}}}
\newcommand{\aaa}{a}
\newcommand{\bbb}{b}

\title{Baryons and Confining Strings}

\author{Oliver Jahn%
  \address[ETH]{Institute for Theoretical Physics, 
    ETH Z\"{u}rich, CH-8093 Z\"{u}rich, Switzerland}%
  \thanks{Talk presented by O. Jahn at Lattice 2003, Tsukuba.}%
  \ and Philippe de Forcrand%
  \addressmark[ETH]%
  \address{CERN, Theory Division, CH-1211 Gen\`{e}ve 23, Switzerland%
}}
       
\begin{document}

\markright{\hfill\rm CERN-TH/2003-217\quad\quad}

\begin{abstract}
  The subleading term of the heavy quark potential (the analogue of
  the L\"{u}scher term) is computed in a string model for the case of
  three quarks.  It turns out to be positive in 2+1 dimensions, making
  the potential non-concave as a function of the scale for fixed
  geometry. The results are compared to numerical simulations of the
  lattice gauge theory.
\vspace{-1pc}
\end{abstract}

\maketitle

\section{Motivation}

The potential of three heavy quarks has recently been the object of
detailed numerical studies \cite{self,Japan}, which lend support to
the so-called Y law, inspired by a string picture of heavy baryons.
Here, we study the leading effect of string fluctuations on the
potential, the analogue of the L\"{u}scher term \cite{LSW,L,LW} in the
quark-antiquark potential.  Understanding such corrections to the Y
law is particularly important since the transition from $\Delta$ to Y
law occurs at large quark separations \cite{self}.

\section{Setup}

We study three static quarks in a $D$-dimensional
Euclidean space with periodic Euclidean time extent $T$.
The potential is obtained from the $T\to\infty$ limit,
\begin{equation}\label{eq:Vqqq}
  V_{\q\q\q} = -\lim_{T\to\infty} \frac1T \ln Z_{\q\q\q}
  \;.
\end{equation}

In the string picture, the classical ground state of the three-quark
system is given by three strings meeting at a junction, whose position
is determined by the requirement of minimal total length of the strings.
The balance of tensions implies angles of $\frac{2\pi}{3}$ between the
strings.  The classical potential is proportional to the total length
of the strings in this configuration,
\begin{equation}\label{eq:Vcl}
  V_{\text{cl}} = \sigma (L_1+L_2+L_3) + m
  = \sigma L_{\textrm{Y}} + m
  \;,
\end{equation}
where $\sigma$ is the same string tension that appears in the mesonic
potential.  $m$ represents the self-energy of the junction.

Following the mesonic case \cite{L}, we shall expand the action to
second order in the transverse fluctuations $\vec\xi_\aaa(t,s)$ of the
string world sheets $\aaa=1,2,3$ around the classical configuration.
Transversality means $\xi_\aaa^D=0$ and $\vec
e_\aaa\cdot\vec\xi_\aaa=0$, where $\vec e_\aaa$ is a spatial unit
vector in the direction of string $\aaa$.  With $\vec\phi(t)$
denoting the position of the fluctuating junction, the boundary
conditions are%
\begin{equation}\label{eq:bc}
  \vec{\xi}_\aaa (t,0) = \vec x_\aaa
  \;, \quad
  \vec{\xi}_\aaa \bigl( t,L_\aaa+\vec e_\aaa{\cdot}\vec\phi(t)
  \bigr) = \vec\phi_{\perp\aaa}(t) 
  \;,
\end{equation}
where $\vec\phi_{\perp\aaa}\equiv\vec\phi-\vec e_\aaa(\vec
e_\aaa\cdot\vec\phi)$, and periodic in $t$.

Invariance under Euclidean transformations inside the plane of the
sheet and perpendicular to it fixes the leading term in a derivative
expansion of the bulk string action to
\begin{equation*}
  S_\fluct^{\text{bulk}} = \frac{\gamma}{2} \sum_{\aaa=1}^3
  \int_{\Gamma_\aaa} \!
  \partial\vec\xi_\aaa \cdot \partial \vec\xi_\aaa 
  \;,
\end{equation*}
where $\Gamma_\aaa$ is the domain of $\xi_\aaa$ implied by the
boundary conditions (\ref{eq:bc}).  We also include a boundary term,
\begin{equation*}
  S_\fluct^{\text{bound}} = \frac{\mu}{2} \int\! \dd t \, |\dot{\vec\phi}|^2
  \;.
\end{equation*}
The change of area caused by fluctuations of the junction in the plane
of a given sheet cancels in the sum over sheets.  The three-quark
potential including leading fluctuation effects can thus be extracted
from the partition function
\begin{equation}\label{eq:Zqqq}
  Z_{\q\q\q} = 
  \int\!\DD\vec\phi
  \int \prod_{\aaa=1}^3 \DD\vec\xi_\aaa \,
  \ee^{-T V_{\text{cl}}-
    S_\fluct^{\text{bulk}} -
    S_\fluct^{\text{bound}}}
  .
\end{equation}

\section{Calculation}

We shall compute $Z_{\q\q\q}$ in two steps.  For fixed $\vec\phi$, each
of the string partition functions,
\begin{equation*}
  Z_\aaa(\vec\phi) = \int\!\DD\xi_\aaa \,
  \exp \left\{ \textstyle
    - \frac{\gamma}{2} \int |\partial\vec\xi_\aaa|^2
  \right\}
  \;,
\end{equation*}
splits into a minimal-area and a fluctuation part,
\begin{equation*}
  Z_\aaa(\vec\phi) = \ee^{ -\frac{\gamma}{2} \int
    |\partial {\vec\xi_{a,\tmin}}|^2 }
  \; \left| \det\nolimits_{\Gamma_\aaa} 
    \bigl( {-\Delta} \bigr) \right|^{-\frac{D-2}{2}}
  \;,
\end{equation*}
where $\vec\xi_{a,\tmin}$ is harmonic and satisfies (\ref{eq:bc}) and
the determinant is computed with Dirichlet boundary conditions on the
domain $\Gamma_\aaa=\{ (t,s) \mid 0\le s\le L_a+\vec
e_\aaa\cdot\vec\phi(t) \}$.

To leading order in $\vec\phi$,
\begin{equation*}
  \vec\xi_{a,\tmin} = \frac{1}{\sqrt{T}} \sum_\omega
  \vec\phi_\omega^\perp \,
  \frac{\sinh(\omega s)}{\sinh(\omega L_\aaa)} \,
  \ee^{\ii\omega t} + \order(\phi^2)
  \;,
\end{equation*}
where $\vec\phi_\omega$ are the Fourier components of $\vec\phi(t)$. 
This implies
\begin{equation*}
  \int |\partial\vec\xi_{a,\tmin}|^2 = \sum_\omega
  \omega \coth(\omega L_\aaa) \, |\vec\phi_\omega^\perp|^2 
  + \order(\phi^3)
  \;,
\end{equation*}
which is the change in minimal area due to $\vec\phi_{\perp\aaa}$.

The determinant induces a renormalisation of both $\sigma$ and $m$.
After applying a Pauli--Villars regularisation, the determinant, which
can be expressed in terms of the heat kernel of $\Delta$, can be
computed by mapping the domain $\Gamma_\aaa$ conformally to a
rectangle $L_\aaa'\times T$.  Since the conformal map cannot change
the ratio $L_\aaa'/T$ (the modular parameter of the cylinder), one
has to choose $L_\aaa'=L_a+\frac1T \int\vec e_\aaa\cdot\vec\phi\,\dd t$.
To leading order in $\vec\phi$, the conformal map is
\begin{equation*}
  f(z) = z + \frac{1}{\sqrt{T}} \sum_{\omega\ne0}
  \frac{\vec e_\aaa\cdot \vec\phi_\omega}{\sinh(\omega L_\aaa)} \,
  \ee^{\omega z} 
  + \order(\phi^2)  
  \;.
\end{equation*}
Following Ref.~\cite{LSW}, the determinant on $\Gamma_\aaa$ can be
related to that on the rectangle, and one finds
\begin{multline*}
  \ln\det\nolimits_{\Gamma_\aaa}(-\Delta)
  = 2 \ln\eta\!\left( \frac{\ii\, T}{2L_\aaa'} \right) 
  \\
  - \frac{1}{12\pi}
  \sum_\omega \omega^3 \coth(\omega L_\aaa) \,
  |\vec e_\aaa{\cdot}\vec\phi_\omega|^2
  \;+\; \order(\phi^3)
  \;,
\end{multline*}
where $\eta$ is Dedekind's function.

The Gaussian integral over $\vec\phi$ in (\ref{eq:Zqqq}) can now be
performed and $V_{\q\q\q}$ be extracted from the limit $T\to\infty$,
cf.\ (\ref{eq:Vqqq}). One obtains:
\begin{align}
  \mskip-36mu
  V_{\q\q\q} &= V_{\text{cl}}^{\text{ren}}
  + V_{1/L}^\parallel + (D{-}3) V_{1/L}^\perp \;
  + \order(L^{-2}) 
  \;, \notag\\
  \mskip-36mu
  V_{1/L}^\parallel &= {-\frac{\pi}{24} \sum_\aaa \frac{1}{L_\aaa}}
  \notag\\
  & \mskip12mu + \int\limits_0^{\infty}\!\frac{\dd\omega}{2\pi} \ln \biggl[
  \tfrac13 \sum_{\aaa<\bbb} 
  {\coth(\omega L_\aaa) \coth(\omega L_\bbb) }
  \biggr] 
  \;, \notag\\
  \mskip-36mu
  V_{1/L}^\perp &= {-\frac{\pi}{24} \sum_\aaa \frac{1}{L_\aaa}}
  \notag\\
  & \mskip12mu + \int\limits_0^{\infty}\!\frac{\dd\omega}{2\pi} \ln \biggl[
  \tfrac13 \sum_{\aaa} { \coth(\omega L_\aaa) }
  \biggr]
  \;. 
  \label{eq:result}
\end{align}
Here, we have separated contributions from fluctuations in the plane
of the three quarks, $V_{1/L}^\parallel$, and perpendicular to it,
$V_{1/L}^\perp$.  Note that both expressions are homogeneous in
$L_\aaa$, representing exactly the $\order(L^{-1})$ term.

\section{Check on the mesonic string}

As a check, we can (artificially) split the string connecting a quark
and an antiquark into two sections of length $L_1$ and $L_2$,
pretending there is a junction in between.  The mass of the junction
should not affect the large-$L$ behaviour, and we should recover the
known result.  In this case, there is no contribution
$V_{1/L}^\parallel$ and the integral in $V_{1/L}^\perp$ becomes
\begin{multline*}
  \int\limits_0^\infty
  \!\frac{\dd\omega}{2\pi} \, \ln \Bigl[ \tfrac12 \bigl( \coth(\omega
  L_1) + \coth(\omega L_2) \bigr) \Bigr] 
  \\
  = \frac{\pi}{24}
  \left( \frac{1}{L_1} + \frac{1}{L_2} - \frac{1}{L_1+L_2}
  \right) 
  \;,
\end{multline*}
which just corrects the L\"{u}scher terms of the two string sections
into that of the full string.

\section{Special cases}

The following special cases reveal interesting consequences of our
result.  In the equilateral case, $L_a=L=\frac{L_{\textrm{Y}}}{3}$, it so happens
that the fluctuations in the plane do not contribute.  Those
perpendicular to it yield
\begin{equation*}
  V_{1/L} = V_{1/L}^\parallel + (D{-}3) V_{1/L}^\perp 
  = - (D-3)\, \frac{\pi}{16 L}
  \;.
\end{equation*}
This means that there is no $1/L$ term in $D=3$.  

\begin{figure}[bthp]
  \vspace*{-9mm}
  \centering
  \noindent\hbox{\includegraphics[height=45mm]{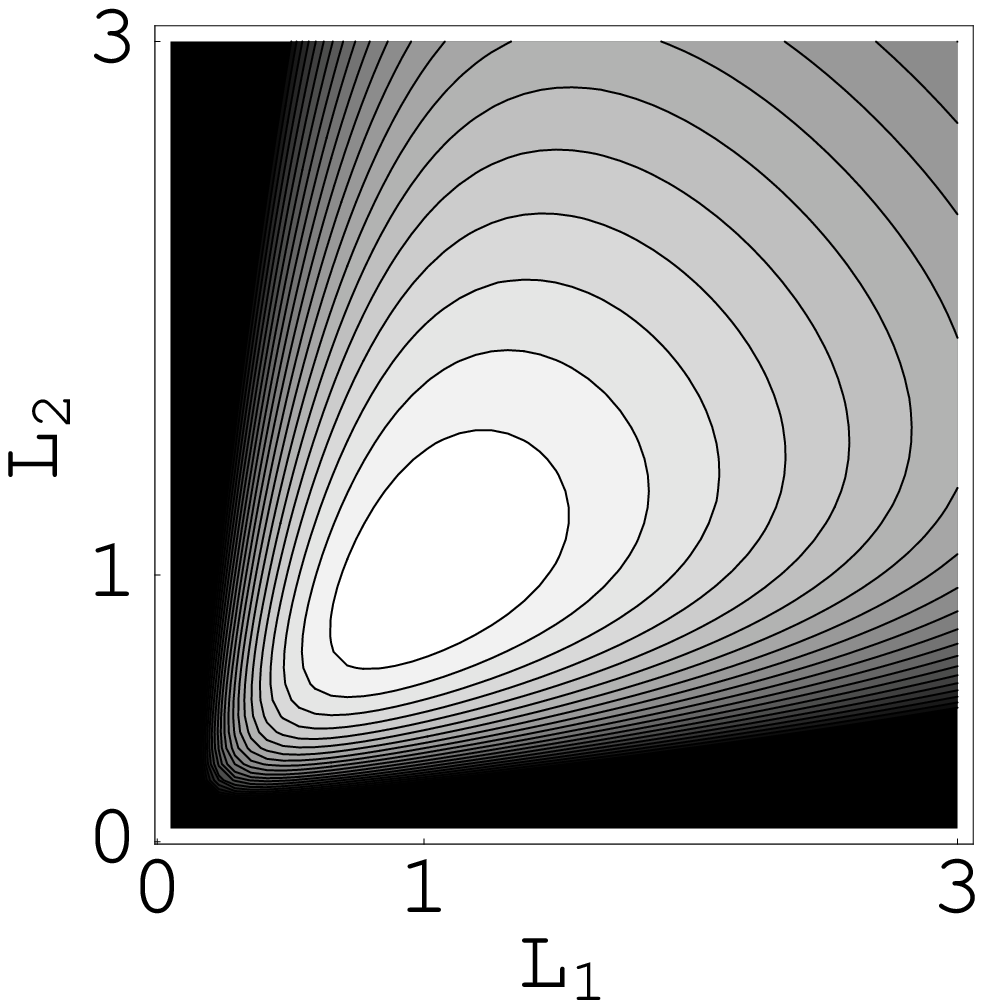}}%
  \ \raisebox{25mm}{$\vcenter{\hbox{\includegraphics[height=45mm]{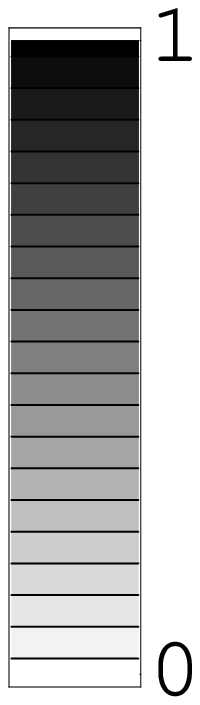}}}%
$}
  \vspace*{-9mm}
  \caption{$\frac{24L_{\textrm{Y}}}{\pi} V_{1/L}(L_1,L_2,1)$ in $D=3$.%
\vspace{-2pc}%
}
  \label{fig:3d}
\end{figure}

Expanding about the equilateral case,
$L_\aaa=(1+\varepsilon_\aaa)L$ with
$\sum_\aaa\varepsilon_\aaa=0$, one finds
\begin{equation*}
  V_{1/L} \approx \frac{\pi}{144 L} \sum_\aaa \varepsilon_\aaa^2
  - (D{-}3) \frac{\pi}{16 L} \biggl(
  1 + \frac{2}{9} \sum_\aaa \varepsilon_\aaa^2
  \biggr)
  \;,
\end{equation*}
so in $D=3$, the $1/L$ term is positive.  This makes $V_{\q\q\q}$
non-concave as a function of the scale $L$ for fixed angles between
the quarks.  This is actually true for all geometries, not just almost
equilateral ones.  Fig.~\ref{fig:3d} shows a density plot of
$V_{1/L}$, as given by Eq.\ (\ref{eq:result}), as a function of $L_1$
and $L_2$ for fixed $L_3=1$.  In $D\ge4$, the $1/L$ term is always
negative, so $V_{\q\q\q}$ is concave.

\section{Lattice gauge theory}

\begin{figure}[tbp]
  \centering
  \vspace*{-2mm}
  \hbox{\includegraphics[width=\hsize,height=.66\hsize]{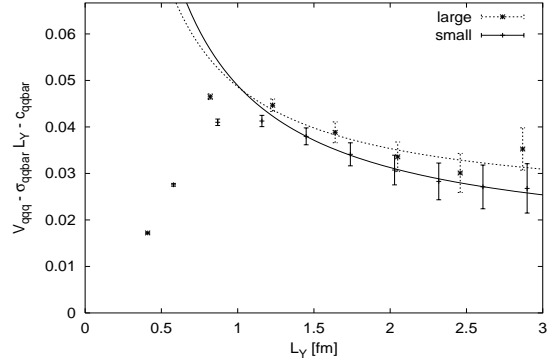}}%
  \vspace*{-8mm}
  \caption{$V_{\q\q\q}-\sigma_{\q\bar\q}L_{\textrm{Y}}-c_{\q\bar\q}$ from lattice simulations 
    for rectangular, isosceles $\q\q\q$ geometries with either the large side 
    or the small sides of the triangle aligned with lattice axes.  The curves are fits $c-b/L_{\textrm{Y}}$.%
The Y law is approached from above.
\vspace{-2pc}%
}
  \label{fig:lattice}
\end{figure}

Numerical simulations of lattice gauge theory were performed in $D=3$.
$V_{\q\q\q}$ was extracted from Polyakov loop correlators using the
method of Ref.~\cite{LW} on a $48^2\times 32$ lattice with lattice
spacing $a\approx0.15\,\text{fm}$ ($\beta=11$), for one qqq geometry.
The deviation from a pure Y law (with the measured mesonic string
tension) is shown in Fig.\ \ref{fig:lattice}.

The asymptotic Y law is clearly approached {\em from above} as predicted, 
making the potential non-concave, but the magnitude
of the $1/L$ term appears too large: $V_{1/L} = 1.2(2) \frac{\pi}{24
  L_{\textrm{Y}}}$ compared to a
coefficient of $0.39$ as predicted by Eq.\ (\ref{eq:result}).  The
discrepancy may be due to the small quark separations: even for the
largest geometries, the shortest string in the classical configuration
is only $0.45\,\text{fm}$ long, so the string picture is hardly
applicable.  In addition, it should be noted that the numerical result is very
sensitive to the estimate of the mesonic string tension.  Finally,
a mandatory extrapolation to $T\to\infty$ has not been performed yet, so
that excited-state contributions may significantly affect the measured
coefficient.

A more thorough check of the string predictions should also include
other geometries.  Finally, it would be very interesting to extend the
calculation to larger $\text{SU}(N)$ gauge groups and compare with
predictions from $1/N$ expansions.

\markright{}


\begin{thebibliography}{9}
\bibitem{self}
C.~Alexandrou, Ph.~de Forcrand and A.~Tsapalis,
Phys.\ Rev.\ D 65 (2002) 054503;
C.~Alexandrou, Ph.~de Forcrand and O.~Jahn,
Nucl.\ Phys.\ Proc.\ Suppl.\ 119 (2002) 649.

\bibitem{Japan}
T.~T.~Takahashi, H.~Matsufuru, Y.~Nemoto and H.~Suganuma,
Phys.\ Rev.\ Lett.\ 86 (2001) 18;
Phys.\ Rev.\ D 65 (2002) 114509.

\bibitem{LSW} M.~L\"{u}scher, K.~Symanzik, P.~Weisz,
  Nucl.\ Phys.\ B 173 (1980) 365.
\bibitem{L} M.~L\"{u}scher, Nucl.\ Phys.\ B 180 (1981) 317.
\bibitem{LW} M.~L\"{u}scher, P.~Weisz, JHEP 0207 (2002) 049.
\end{thebibliography}
\end{document}